\documentclass[aps,prd,preprint,showpacs]{revtex4}
\usepackage{graphicx}
\usepackage{subfigure}
\usepackage{hyperref}
\begin{document}
\draft
\title{Anomalous single production of fourth generation $t'$ quarks at ILC and CLIC}
\author{A. Senol}
\email{asenol@kastamonu.edu.tr}
\author{A.~T. Tasci}
\email{atasci@kastamonu.edu.tr}
\author{F. Ustabas}
\email{fatmaustbs@gmail.com}
 \affiliation{Department of
Physics, Kastamonu University, 37100, Kastamonu, Turkey}

\begin{abstract}
We present a detailed study of the anomalous single fourth
generation $t'$ quark production within the dominant Standard Model
(SM) decay modes at future $e^+e^-$ colliders. We calculate the
signal and background cross sections in the mass range 300-800 GeV.
We also discuss the limits of $t'q\gamma$ and $t'qZ$ ($q=u,c$)
anomalous couplings as well as values of attainable integrated
luminosity for 3$\sigma$ observation limit.
\end{abstract}
\maketitle
\section{introduction}
As it is well known that Standard Model (SM) of particle physics
includes three generation of fermions, but does not rule out a
fourth generation. The only restriction of the fermion generation
numbers comes from asymptotic freedom constraint in QCD is less than
nine. On the other hand, the bounds from reanalyzed electroweak
precision measurements has shown that existence of fourth fermion
generations is not prohibited \cite{He:2001tp,Kribs:2007nz}.
Therefore, the SM can be simply extended with a sequential
repetition as four quark and four lepton left handed doublets and
corresponding right handed singlets.

A possible fourth generation may play an important role in
understanding the well-known unanswered question such as the CP
violation and flavor structure of standard theory
\cite{Hou:2010,BarShalom:2009sk,Buras:2010pi,Soni:2008bc,Eberhardt:2010bm,Soni:2010xh,Alok:2010zj},
electroweak symmetry breaking \cite{Holdom:1986rn,
Hill:1990ge,Elliott:1992xg,Hung:2010xh}, hierarchies of fermion mass
spectrum and mixing angle in quark/lepton sectors
\cite{Holdom:2006mr,Hung:2007ak,Hung:2009ia,Hung:2009hy,Cakir:2009xi}.
Also, the theoretical and experimental aspects of fourth SM
generation in a recent review can be found in Ref.
\cite{Holdom:2009rf}.

So far, the current 95 \% C.L. mass lower limits of the fourth
generation quarks from experimental measurements at the Tevatron are
$m_{t'}> 311$ GeV \cite{Lister:2008is} and $m_{b'}> 338$ GeV
\cite{Aaltonen:2009nr}, whereas the partial wave unitarity gives an
upper bound of about 600 GeV \cite{Chanowitz:1978mv}.

The determination of allowed parameter space of fourth generation
fermions will be an important goal of the LHC era. Masses of the
fourth generation quarks will already be well known from the LHC
data before a TeV scale linear collider runs, since (if they exist)
they are predicted to be produced in pairs at the LHC \cite{atlas}.
The large value of their masses would provide special advantage to
new interactions originating at a higher scale. The precise
determination of fourth generation quark properties may present the
existence of physics beyond the SM. The fourth generation quarks can
couple to the gauge bosons via new interactions as well as the
flavor changing neutral current (FCNC) interactions with anomalous
couplings as $t'qV~(V=\gamma,~Z,~g;~ q=u,c)$. The FCNC vertices can
be described by an effective Lagrangian which contains series in
power of $\kappa/\Lambda$ where $\kappa$ is the anomalous magnetic
moment type coupling and $\Lambda$ is the cut off scale of new
interactions. Nevertheless, the LHC may not provide us with
sufficient information about some parameters of the fourth
generation quarks. But, a linear collider with energies on the TeV
scale, extremely high luminosity and clean experimental environment,
can provide complementary information for these parameters with
performing precision measurements that would complete the LHC
results. Most popular proposed linear colliders with energies on the
TeV scale and extremely high luminosity are International Linear
Collider (ILC) \cite{:2007sg} and Compact Linear Collider (CLIC)
\cite{Braun:2008zzb}. Due to the anomalous interactions serious
contributions can be expected for the production of the fourth
generation fermions. These anomalous effects of the fourth
generation quarks has been studied in phenomenological perspective
at hadron colliders
\cite{Arik:2003vn,Arik:2002sg,Cakir:2009ib,Ciftci:2008tc,Sahin:2010wg}
and also at future ep colliders \cite{Alan:2003za,Ciftci:2009it}. In
this study, we investigate the single production of fourth
generation $t'$ quarks at proposed linear colliders via anomalous
interactions. After its production, we assume the dominant SM $t'$
decay channel as $t'\to W b$. The aim of this study is to find
discovery potential of $t'$ parameters from a detailed analysis for
signal and background including Monte Carlo simulation. Thus, we
implement the related interaction vertices into the CompHEP package
\cite{Pukhov:1999gg} and study the FCNC parameters in detail as well
as the effects of initial state radiation (ISR) and beamstrahlung
(BS) in the $e^+e^-$ collisions.
\section{Interactions of fourth generation $t'$ quarks}
Before examining the single productions of the $t'$ quarks, an
extension the SM with fourth generation is needed. One of the
approach for an extension of the SM is simply to add a sequential
fourth family fermions where left-handed components transform as a
doublet of $SU(2)_{L}$ and right-handed components as singlets. The
interaction of fourth generation $t'$ quarks with three known
generation of quarks, $q_i$, via  the SM gauge bosons
($\gamma,g,Z^{0},W^{\pm}$) is given by
\begin{eqnarray}
L_s & = & -g_{e}Q_{t'}\overline{t}'\gamma^{\mu}t'A_{\mu}\nonumber \\
 &  & -g_{s}\overline{t}'T^{a}\gamma^{\mu}t'G_{\mu}^{a}\nonumber \\
 &  & -\frac{g}{2c_{W}}\overline{t}'\gamma^{\mu}(g_{V}-g_{A}\gamma^{5})t'Z_{\mu}^{0}\nonumber \\
 &  & -\frac{g}{2\sqrt{2}}V_{t'q_{i}}\overline{t}'\gamma^{\mu}(1-\gamma^{5})q_{i}W_{\mu}^{\pm}+h.c.\label{eq:1}
 \end{eqnarray}
where $g_{e}$, $g$ are the electro-weak coupling constants, and
$g_{s}$ is the strong coupling constant. $A_{\mu}$, $G_{\mu}$,
$Z_{\mu}^{0}$ and $W_{\mu}^{\pm}$ are the vector fields for photon,
gluon, $Z^{0}$-boson and $W^{\pm}$-boson, respectively. $Q_{t'}$ is
the electric charge of fourth family quark $t'$; $T^{a}$ are the
Gell-Mann matrices. $g_{V}$ and $g_{A}$ are the vector and
axial-vector type couplings of the neutral weak current with $t'$
quark, $\theta_W$ is the weak mixing angle and $c_W=\cos\theta_W$.
Finally, the $V_{t'q}$ denotes the elements of extended 4$\times$4
CKM mixing matrix which are constrained by flavor physics. In this
study, we use the parametrization of extended CKM matrix elements,
$V_{t'd}$=0.0044, $V_{t's}$=0.114 and $V_{t'b}$=0.22
\cite{Hou:2006mx}. We calculate the decay width of $t'$ via $t'\to W
q$ process as
\begin{eqnarray}
\Gamma(t'\to W
q)=\frac{\alpha_e^2|V_{t'q_i}|^2}{16s_W^2m_{t'}^3M_W^2}\{m_{t'}^6-3m_{t'}^2M_W^4+2M_W^6\}
\end{eqnarray}
where $s_W=\sin\theta_W$, $g=g_e/s_W$, $g_e=\sqrt{4\pi\alpha_e}$ and
$M_W$ is the mass of $W$ boson.

In the SM, FCNC interactions are absent at tree level. However, top
quark sector is suitable for searching anomalous FCNC interactions
as being heaviest particle to date discovered
\cite{Fritzsch:1999rd}. The fourth generation $t'$ quarks which are
heavier than the top quark can also couple to the FCNC currents. The
effective Lagrangian for the anomalous magnetic moment type
interactions among the fourth family quarks $t'$, ordinary quarks
$q$, and the neutral gauge bosons $V=\gamma,Z,g$ are given by
\begin{eqnarray}
L'_{a} & = & \sum_{q_{i}=u,c,t}\frac{\kappa_{\gamma}^{q_{i}}}{\Lambda}Q_{q_{i}}g_{e}\overline{t}'\sigma_{\mu\nu}q_{i}F^{\mu\nu}+\sum_{q_{i}=u,c,t}
\frac{\kappa_{Z}^{q_{i}}}{2\Lambda}\frac{g}{c_W}\overline{t}'\sigma_{\mu\nu}q_{i}Z^{\mu\nu}\nonumber \\
 &  & +\sum_{q_{i}=u,c,t}\frac{\kappa_{g}^{q_{i}}}{2\Lambda}g_{s}\overline{t}'\sigma_{\mu\nu}\lambda_{a}q_{i}G_{a}^{\mu\nu}+h.c.
 \end{eqnarray}
where $F^{\mu\nu}$, $Z^{\mu\nu}$ and $G^{\mu\nu}$ are the field
strength tensors of the gauge bosons;
$\sigma_{\mu\nu}=i(\gamma_{\mu}\gamma_{\nu}-\gamma_{\nu}\gamma_{\mu})/2$;
$\lambda_{a}$ are the Gell-Mann matrices; $Q_{q_i}$ is the electric
charge of the quark ($q$). $\kappa_{\gamma}$, $\kappa_{Z}$ and
$\kappa_{g}$ are the anomalous coupling with photon, $Z$ boson and
gluon, respectively. $\Lambda$ is the cut-off scale for the new
interactions.

Anomalous decay widths of $t'$ quarks which are calculated by using
the effective Lagrangian are given below
\begin{eqnarray}
\Gamma(t'\to
q\gamma)&=&\frac{\kappa_{\gamma}^2}{2\Lambda^2}\alpha_eQ_{q_i}^2m_{t'}^3\\
\Gamma(t'\to q
g)&=&\frac{\kappa_{g}^2}{3\Lambda^2}\alpha_sm_{t'}^3\\
\Gamma(t'\to q
Z)&=&\frac{\kappa_{Z}^2\alpha_e}{16m_{t'}^3\Lambda^2s_W^2c_W^2}(2m_{t'}^6-3M_Z^2m_{t'}^4+M_Z^6)
\end{eqnarray}
where $M_Z$ is the mass of the $Z$ boson. All numerical calculations
have been performed with CompHEP \cite{Pukhov:1999gg} by including
the new interaction vertices. In Table~\ref{tab1}, we give the
numerical values of the total decay widths and branching ratios for
all decay channels of $t'$ quarks by
$\kappa_{\gamma}/\Lambda=\kappa_{Z}/\Lambda=\kappa_{g}/\Lambda=0.1~
$TeV$^{-1}$
\begin{figure}[hptb!]
\includegraphics[width=6cm]{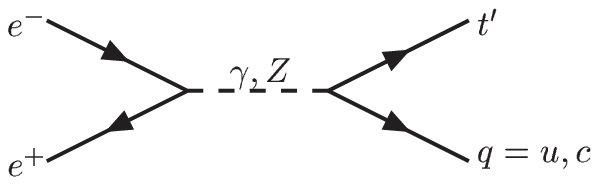}
\caption{Feynman diagram for single production of $t'$ in $e^+e^-$
collision.}\label{fig1}
\end{figure}

The contributing Feynman Diagrams for anomalous single $t'$ produced
in $e^+e^-$ collision are shown in Fig.~\ref{fig1}. The analytical
expression for cross section of $e^+e^-\to t'q (q=u,c)$ process,
calculated by using Lagrangian $L_a$, is found as

\begin{eqnarray}
\sigma&=&\frac{\pi\alpha_e^2(2m_{t'}^6-3m_{t'}^4s+s^3)}{48s_W^4c_W^4\Lambda^2s^3[(s-M_Z^2)^2+\Gamma_Z^2M_Z^2]}
[9s^2(1-4s_W^2+8s_W^4)
\kappa_Z^2\nonumber\\&+&48s_W^2c_W^2\kappa_Z\kappa_{\gamma}(1-4s_W^2)s(s-M_Z^2)+128s_W^4c_W^4\kappa_{\gamma}^2((s-M_Z^2)^2+\Gamma_Z^2M_Z^2)]
\end{eqnarray}
where $\Gamma_Z$ is the total decay width of $Z$ boson.
\begin{table}
\caption{Total decay widths (GeV) and branching ratios (\%) of the
$t'$ quarks in both chiral and anomalous interactions.}\label{tab1}
\begin{tabular}{lcccccccccccccccccccr}
  \hline
  $m_{t'}$(GeV) && $Wd$ && $Wb$ && $Ws$&&$Zu(c)$&&$Zt$&&$gu(c)$&&$gt$&&$\gamma u(c)$&&$\gamma t$&&$\Gamma_{tot}$(GeV)
  \\\hline
  300 &&0.029 &&72&&19&&0.22&&0.024&&3.6&&1.0&&0.079&&0.023&&0.59 \\
  400 &&0.029 &&71&&19&&0.23&&0.10&&3.5&&1.9&&0.078&&0.041&&1.43 \\
  500 &&0.028 &&71&&19&&0.23&&0.15&&3.5&&2.3&&0.077&&0.052&&2.82 \\
  600 &&0.028 &&71&&19&&0.23&&0.17&&3.5&&2.6&&0.077&&0.059&&4.89 \\
  700 &&0.028 &&70&&19&&0.24&&0.19&&3.4&&2.8&&0.077&&0.063&&7.79 \\
  800 &&0.028 &&70&&19&&0.24&&0.20&&3.4&&3.0&&0.076&&0.066&&11.6 \\
  \hline
  \end{tabular}
  \end{table}
\section{Signal and Background Analysis}
\begin{figure}[hptb!]
 \subfigure[]{\includegraphics[width=8cm]{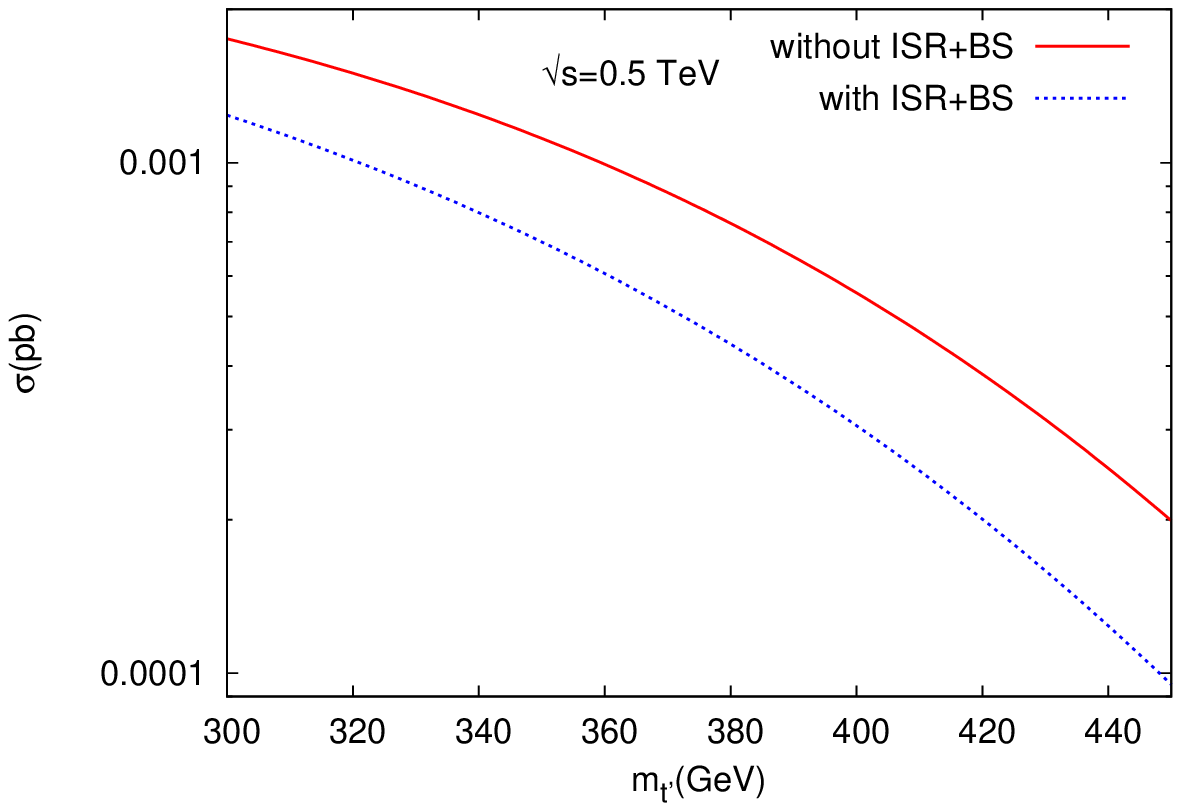}}
\hspace{0.1cm}
\subfigure[]{\includegraphics[width=8cm]{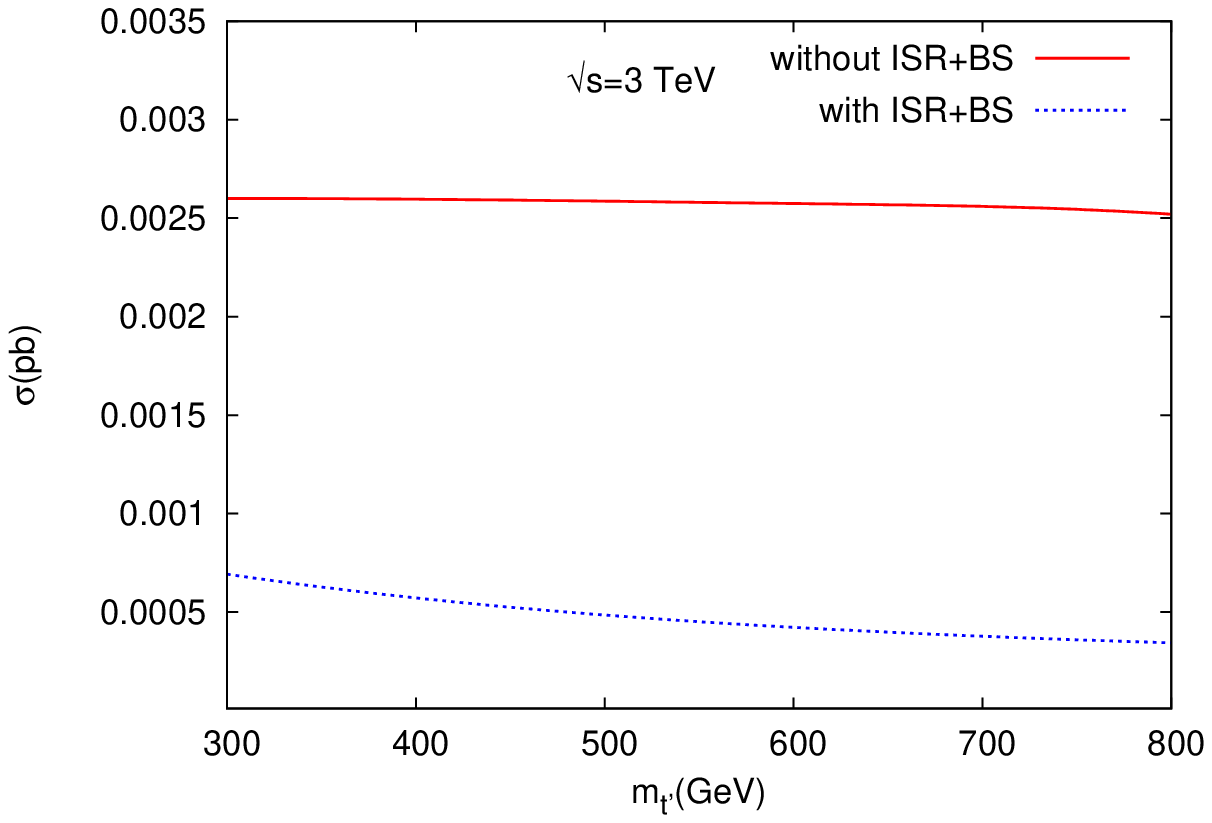}} \caption{The
total cross sections in pb for the process $e^+e^-\to t'q (q=u,c)$,
as function of ${m_t'}$ at (a) $\sqrt{s}=$0.5 TeV and (b)
$\sqrt{s}=$3 TeV.}\label{fig:cs1}
\end{figure}

The total cross sections for single production of $t'$ quarks are
plotted in Fig.~\ref{fig:cs1} with respect to their masses at
collision center of mass energies (a) 0.5 TeV and (b) 3 TeV with
assumption $\kappa/\Lambda=0.1$ TeV$^{-1}$. A specific feature of
the linear colliders is the occurrence of initial state radiation
(ISR) and beamstrahlung (BS). When calculating the ISR and BS
effects, we take beam parameters which are given in Table~\ref{tab2}
for the ILC and CLIC. In Fig.~\ref{fig:cs1}, the dotted line denotes
the cross section including ISR+BS effects while solid line denotes
without ISR+BS effects. The effects of ISR and BS can lead to a
decrease in the cross section. After this point, we take into
account ISR+BS effects in all our numerical calculations.
\begin{table}
\caption{Main parameters of ILC and CLIC. Here N is the number of
particles in bunch. $\sigma_{x}$ and $\sigma_{y}$ are beam sizes,
$\sigma_{z}$ is the bunch length.}\label{tab2}
\begin{tabular}{lcc}
\hline Parameters  & ILC   & CLIC\tabularnewline \hline
$E_{cm}(\sqrt{s})$ TeV  & $0.5$   & $3$\tabularnewline
$L(10^{34}cm^{-2}s^{-1})$  & $2$   & $5.9$\tabularnewline
$N$$(10^{10})$  & $2$  & $0.372$\tabularnewline  $\sigma_{x}$ (nm) &
$640$   & $45$\tabularnewline  $\sigma_{y}$ (nm)  & $5.7$ &
$1$\tabularnewline  $\sigma_{z}$ ($\mu$m)  & $300$ &
$44$\tabularnewline \hline
\end{tabular}

\end{table}
The single production of fourth generation $t'$ quarks of signal
process including dominance of the SM decay mode over anomalous
decay is
\begin{eqnarray}
e^+e^-\longrightarrow t' q_i\longrightarrow W^+ q_j q_i ~~~~~~
(q_i=u,c ;~~ q_j=d,s,b~~)
\end{eqnarray}
\begin{figure}[hptb!]
 \subfigure[]{\includegraphics[width=8cm]{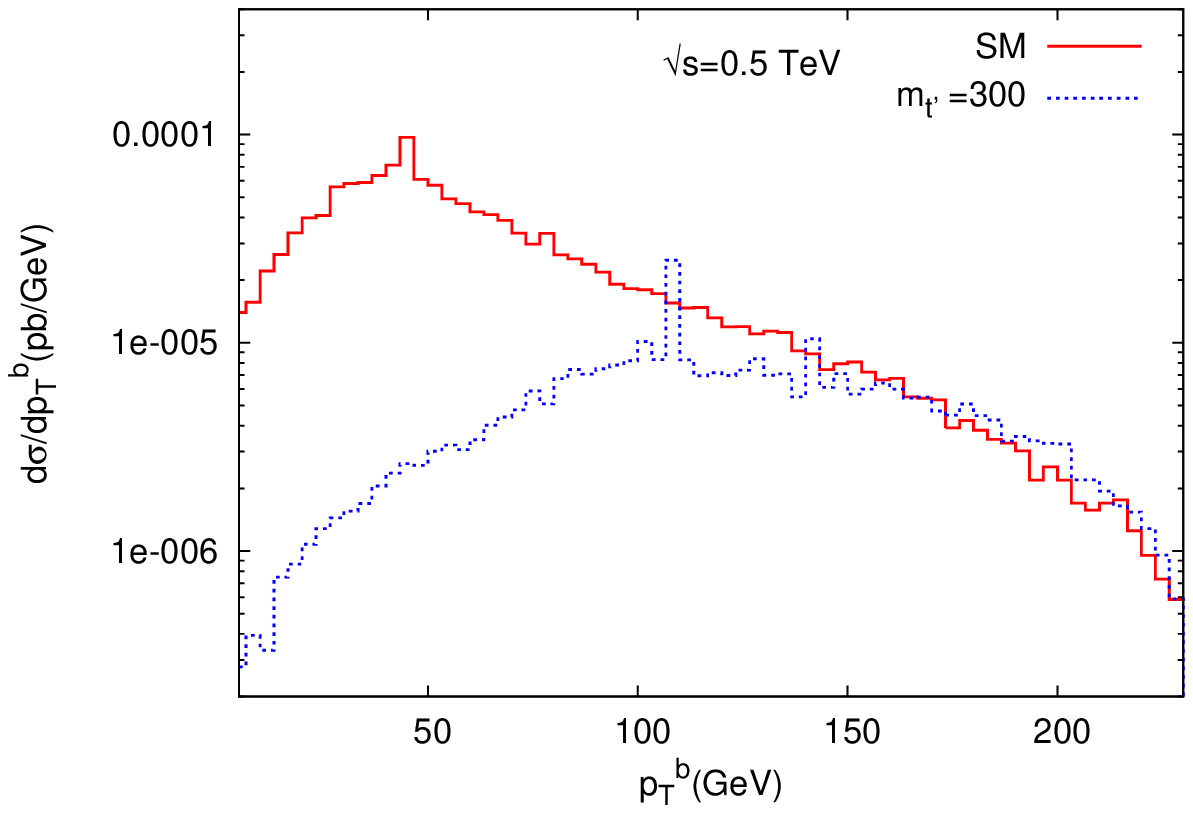}}
\hspace{0.1cm}
\subfigure[]{\includegraphics[width=8cm]{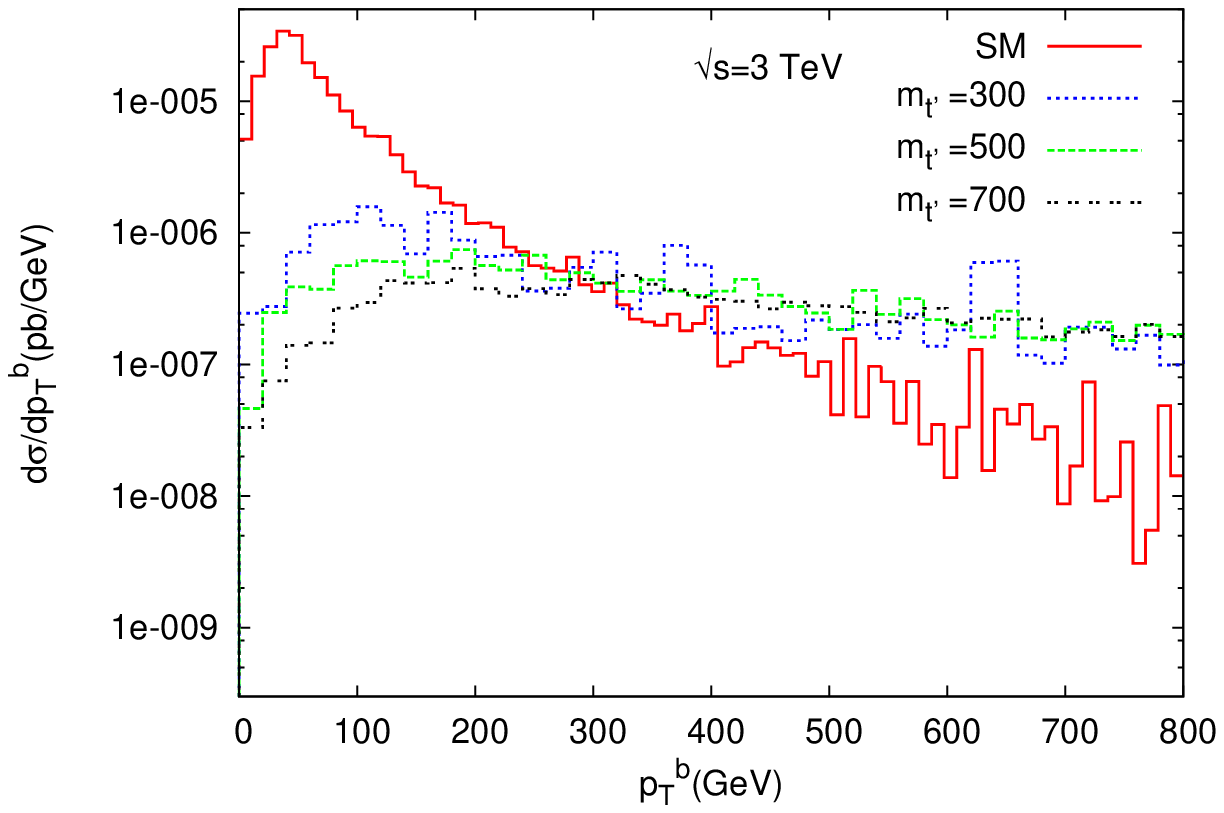}} \caption{The
differential cross section depending on the transverse momentum of
the final state $b$ quark of process $e^+e^-\to W^{+}bq (q=u,c)$ for
SM background (solid line) and signal with different mass values of
$t'$ quarks at (a) $\sqrt{s}=$0.5 TeV and (b) $\sqrt{s}=$3
TeV.}\label{fig2}
\end{figure}
The transverse momentum ($p_T$) distributions of the final state
 $b$ quark for signal and background are shown in Fig.~\ref{fig2} for ILC and CLIC options. Comparing
 the signal $p_T$ distribution of $b$ quark with that of the corresponding
 background, we applied a $p_T$ cut of $p_T> 50$ GeV to reduce the background.

\begin{figure}[hptb!]
 \subfigure[]{\includegraphics[width=8cm]{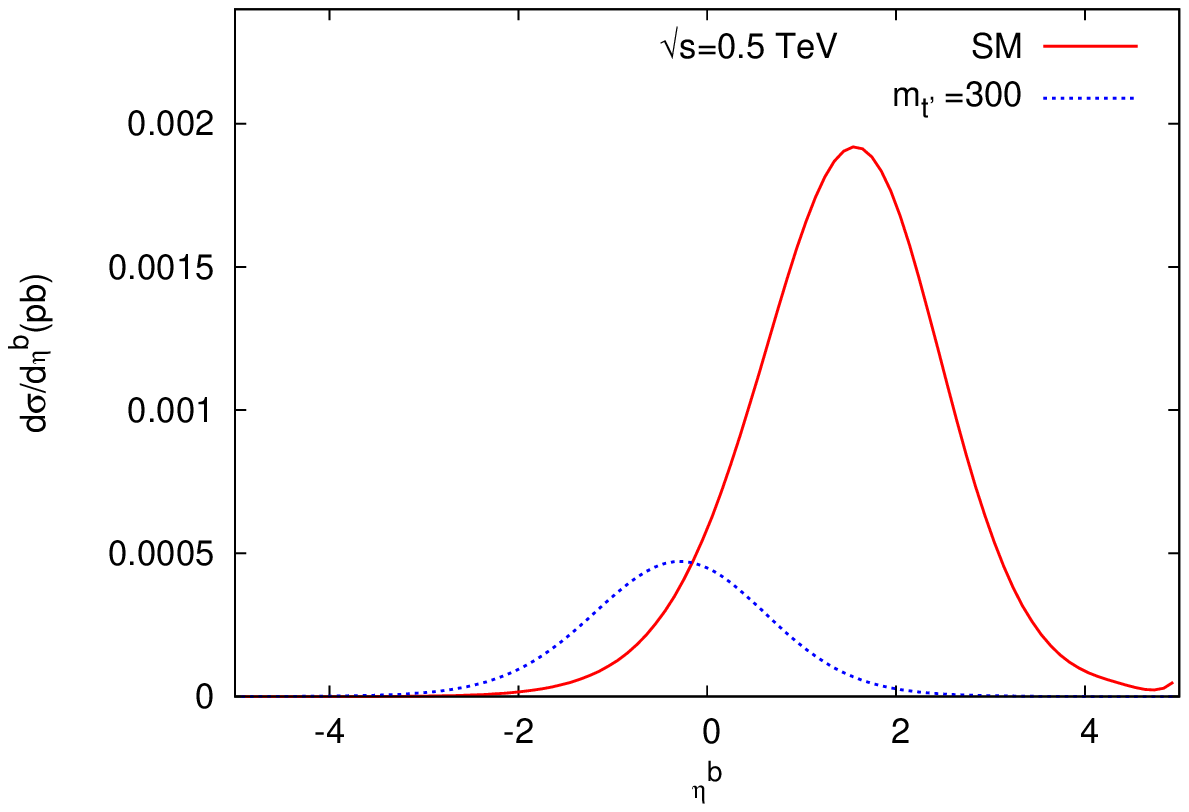}}
\hspace{0.1cm}
\subfigure[]{\includegraphics[width=8cm]{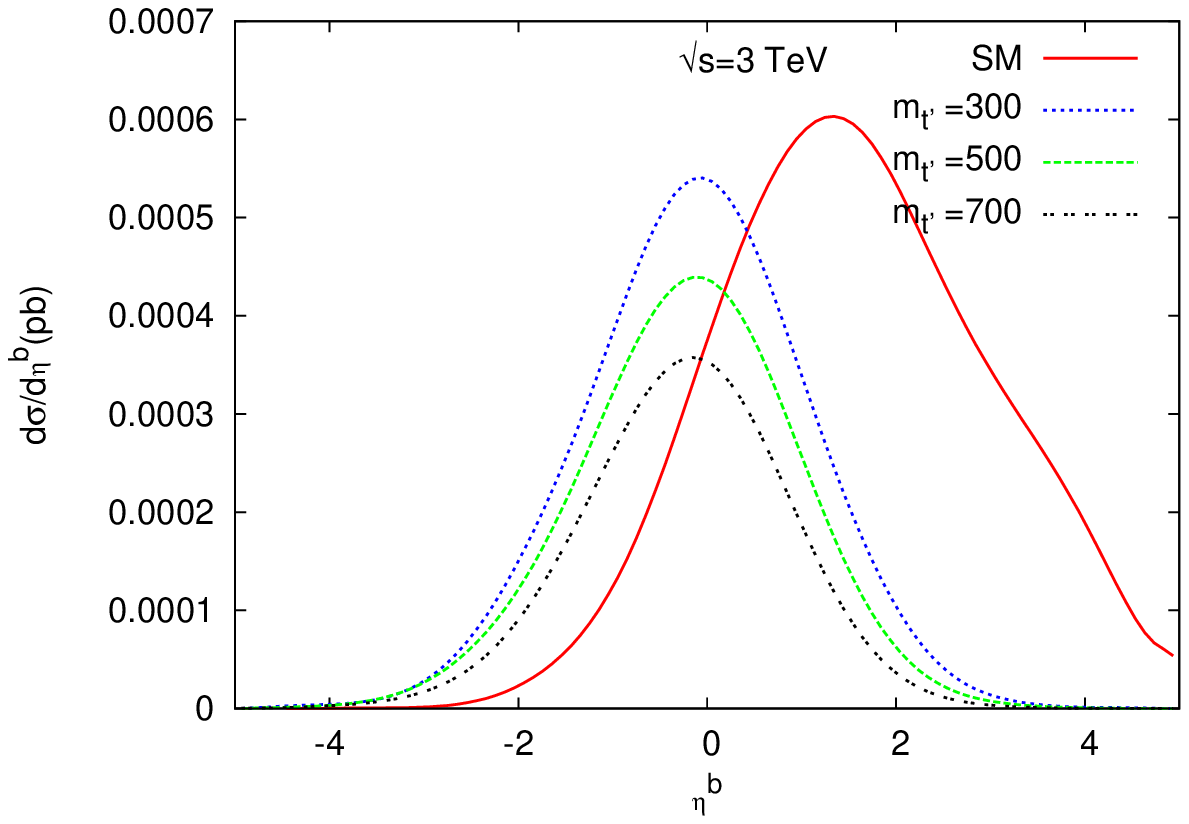}} \caption{The
rapidity distribution of the final state $b$ quark of the process
$e^+e^-\to W^+bq (q=u,c)$ at (a) $\sqrt{s}=$0.5 TeV and (b)
$\sqrt{s}=$3 TeV.}\label{fig3}
\end{figure}

\begin{figure}[hptb!]
\subfigure[]{\includegraphics[width=8cm]{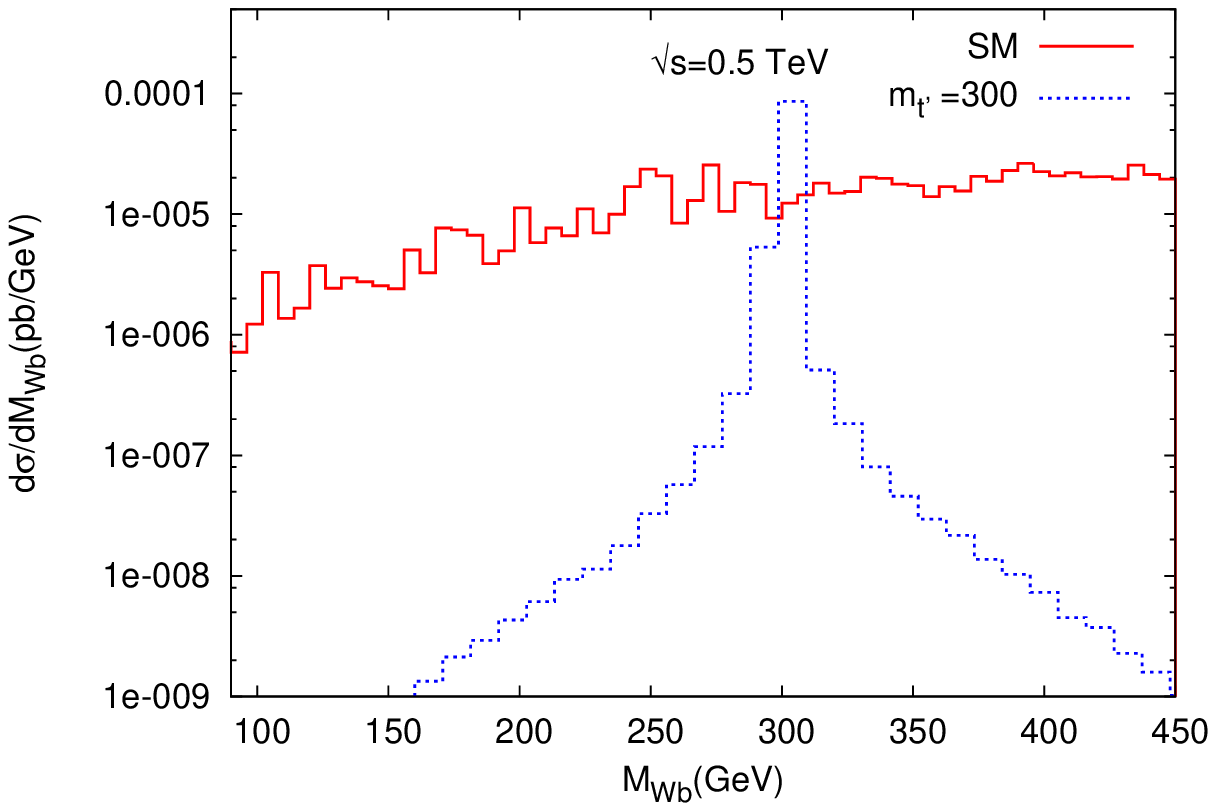}}
\hspace{0.1cm}
\subfigure[]{\includegraphics[width=8cm]{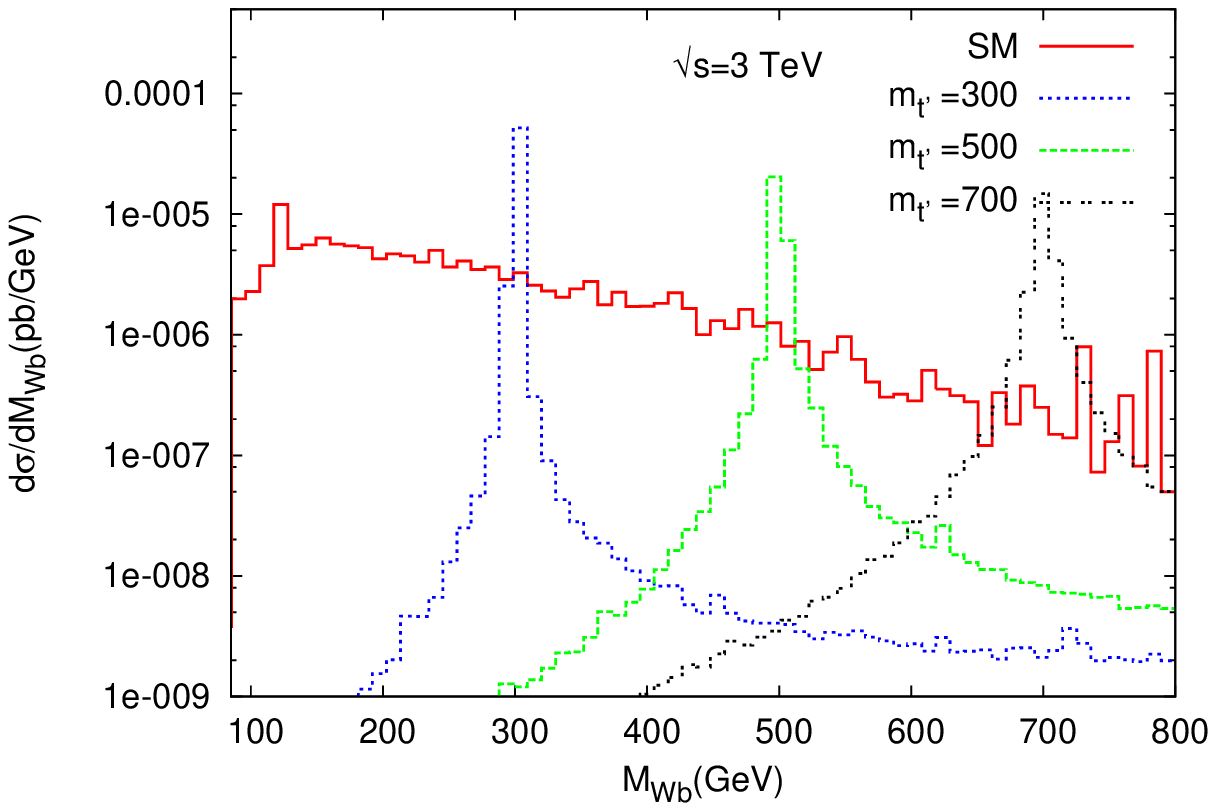}}
\caption{The invariant mass distribution of the final state $Wb$
system for SM background (solid line) and signal from $t'$ decay for
(a) $m_{t'}=300$ GeV (dotted line) at $\sqrt{s}=$0.5 TeV and (b)
$m_{t'}=300$ GeV (dotted line),  $m_{t'}=500$ GeV (dashed line) and
$m_{t'}=700$ GeV (dot-dashed line) at $\sqrt{s}=$3 TeV.}\label{fig4}
\end{figure}

 The rapidity distributions of final state $b$ quark in signal and
 background processes are plotted in Fig.~\ref{fig3}. From these figures, we can
 see that the cut $|\eta^b|<2$ can be applied to suppress the
 background while the signal remains almost unchanged.
 We plot the invariant mass distributions for the $W^+b$ system in
 the final state, the signal has a peak around mass of $t'$ quark
 over the background as shown in Fig.~\ref{fig4}.

\begin{figure}[hptb!]
 \subfigure[]{\includegraphics[width=8cm]{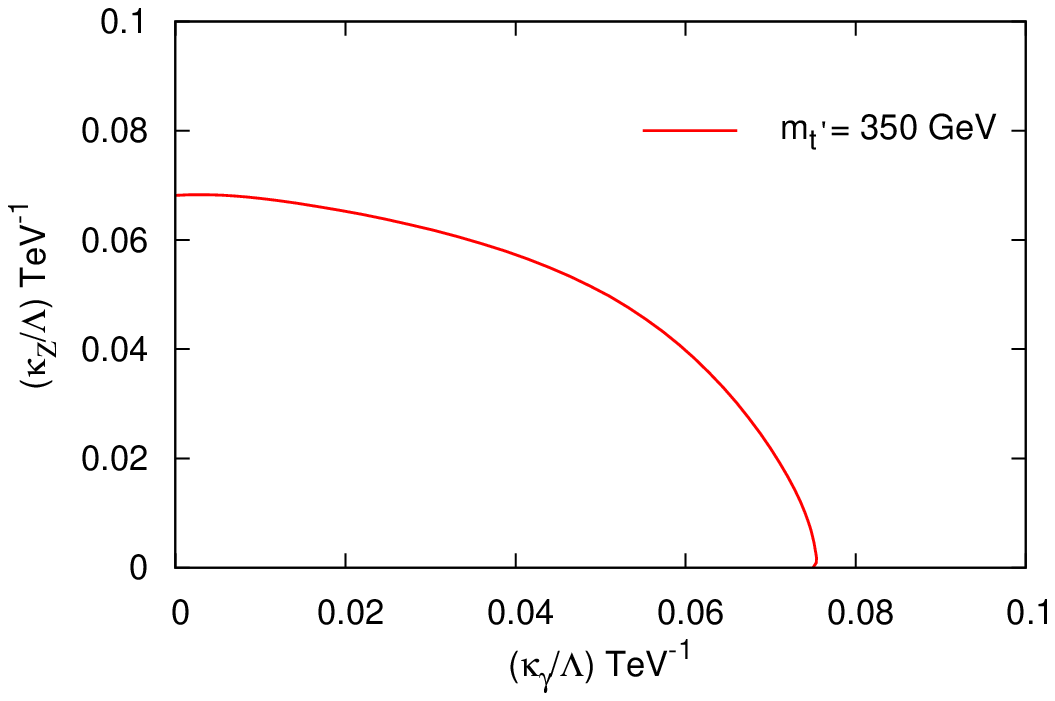}}
\hspace{0.1cm}
\subfigure[]{\includegraphics[width=8cm]{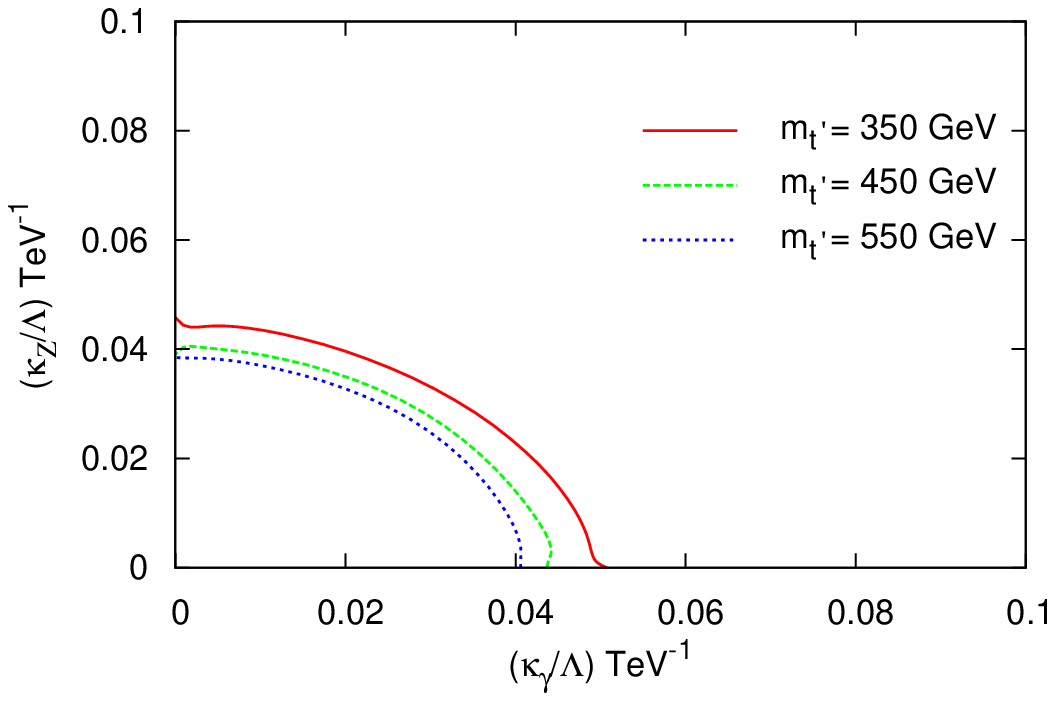}}
\caption{The 3$\sigma$ contour plot for the anomalous couplings
reachable at (a) $\sqrt{s}=$0.5 TeV with $L_{int}$=2$\times10^5$
pb$^{-1}$ and (b) $\sqrt{s}=$3 TeV with $L_{int}$=5.9$\times10^5$
pb$^{-1}$.}\label{cp1}
\end{figure}

\begin{figure}[hptb!]
\includegraphics[width=11cm]{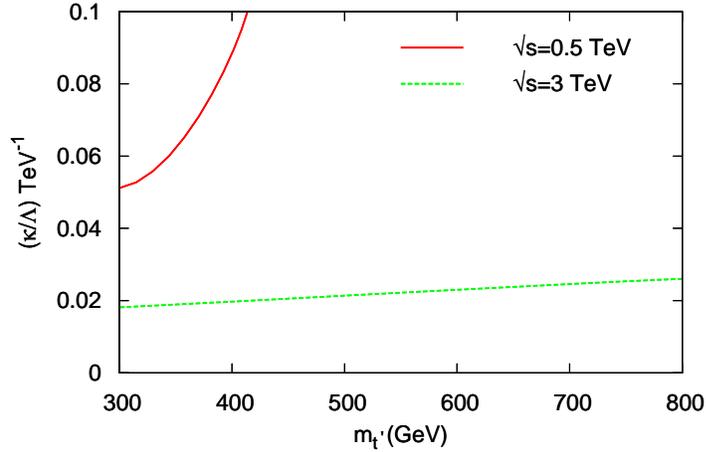}
\caption{The 3$\sigma$ contour plot for the discovery of $t'$ quarks
at the center of mass energies of ILC and CLIC with luminosities of
2$\times10^5$ pb$^{-1}$ and 5.9$\times10^5$ pb$^{-1}$.  }\label{cp2}
\end{figure}

In order to discuss the observability of the fourth generation $t'$
quarks at linear colliders, we need to calculate signal and
background cross sections as well as the statistical significance
($SS$) which are given in Tables~\ref{ss05} and \ref{ss3} for ILC
and CLIC parameters, respectively. We obtain the $SS$  of the signal
by using the formula \cite{Ball:2007zza},
\begin{eqnarray*}
SS&=&\sqrt{2L_{int}\epsilon[(\sigma_{S}+\sigma_{B})\ln(1+\sigma_{S}/\sigma_{B})-\sigma_{S}]}.
\end{eqnarray*}
where $\sigma_{S}$ and $\sigma_{B}$ are the signal and background
cross sections, respectively. For realistic analysis we take into
account finite energy resolution of the detectors. We use the mass
bin width $\Delta m=max(2\Gamma,\delta m)$ in our numerical
calculations to count signal and background events with the mass
resolution $\delta m$. We also apply the mentioned $p_T$ and $\eta$
cuts assuming the integrated luminosities given in Table~\ref{tab2}.
Here, we also consider the final state $W$ boson in the signal and
background processes decay leptonicaly via $W^{\pm}\to l^{\pm}\nu_l$
where $l^{\pm}=e^{\pm},\mu^{\pm}$ and we assume the b-tagging
efficiency as $\epsilon= 50\%$. From Tables~\ref{ss05} and
\ref{ss3}, we see that $t'$ quarks can be observed in the scanned
mass interval of 300-800 GeV at the CLIC while up to about 400 GeV
at the ILC with the observability criteria of 3$\sigma$ by taking
the anomalous couplings
$\kappa_{\gamma}/\Lambda=\kappa_{Z}/\Lambda=\kappa_{g}/\Lambda=0.1~
$TeV$^{-1}$.
\begin{table}
\caption{The signal and background cross sections in pb and signal
 Statistical Significance ($SS$) for the ILC at $\sqrt s=0.5$ TeV with integrated
 luminosity of 2$\times10^5$ pb$^{-1}$ by assuming $\kappa_{\gamma}/\Lambda=\kappa_{Z}/\Lambda=\kappa_{g}/\Lambda=0.1
~$TeV$^{-1}$.}\label{ss05}
\begin{tabular}{lccc}
  \hline
  $m_{t'}$(GeV) & $\sigma_{S}$(pb) & $\sigma_{B}$(pb) & SS \\\hline
  300 & $1.17\times10^{-3}$ & $5.08\times10^{-5}$ & 10.92 \\
  350 & $7.88\times10^{-4}$ & $9.69\times10^{-5}$ & 7.17\\
  400 & $3.95\times10^{-4}$ & $1.58\times10^{-4}$ & 3.62 \\
  450 & $9.09\times10^{-5}$ & $2.05\times10^{-4}$ & 0.88 \\
  \hline
\end{tabular}
\end{table}
\begin{table}
\caption{The signal and background cross sections in pb and signal
 Statistical Significance ($SS$) for the CLIC at $\sqrt s=3$ TeV with integrated
 luminosity of 5.9$\times10^5$ pb$^{-1}$ by assuming $\kappa_{\gamma}/\Lambda=\kappa_{Z}/\Lambda=\kappa_{g}/\Lambda=0.1
~$TeV$^{-1}$.}\label{ss3}
\begin{tabular}{lccccc}
  \hline
  $m_{t'}$(GeV) & $\sigma_{S}$(pb)& & $\sigma_{B}$(pb) && SS \\\hline
  300 & $1.48\times10^{-3}$ && $2.01\times10^{-5}$ && 25.68 \\
  400 & $1.27\times10^{-3}$ && $1.88\times10^{-5}$ && 23.47 \\
  500 & $1.05\times10^{-3}$ && $1.77\times10^{-5}$ && 20.96 \\
  600 & $1.09\times10^{-3}$ && $2.32\times10^{-5}$ && 20.62 \\
  700 & $9.55\times10^{-4}$ && $2.77\times10^{-5}$ && 18.35 \\
  800 & $8.40\times10^{-4}$ && $3.17\times10^{-5}$ && 16.44 \\
  \hline
\end{tabular}
\end{table}
Up to now, we assume the anomalous couplings are equal to each
other. To analyze the case of
$\kappa_{\gamma}/\Lambda\neq\kappa_Z/\Lambda$, the 3$\sigma$ contour
plots for the anomalous couplings in the
$\kappa_{\gamma}/\Lambda-\kappa_Z/\Lambda$ plane are presented in
Fig.~\ref{cp1} a) for $\sqrt s=0.5$ TeV and b) for $\sqrt s=3$ TeV
with different mass values of $t'$ quarks. According to these
figures the lower limits of $\kappa_{\gamma}/\Lambda$ and
$\kappa_Z/\Lambda$ are about 0.07 TeV$^{-1}$ at the ILC and 0.05
TeV$^{-1}$ at the CLIC energies for $m_{t'}$=350 GeV. We plot the
3$\sigma$ contours in the $\kappa/\Lambda$-$m_{t'}$ plane for
different masses of $t'$ quarks as shown in Fig.~\ref{cp2}. The area
of the allowed parameter space of $t'$ quarks are above the lines in
Figs.~\ref{cp1} and \ref{cp2} by taking into account the extended
CKM matrix elements.

\begin{figure}[hptb!]
\subfigure[]{\includegraphics[width=8cm]{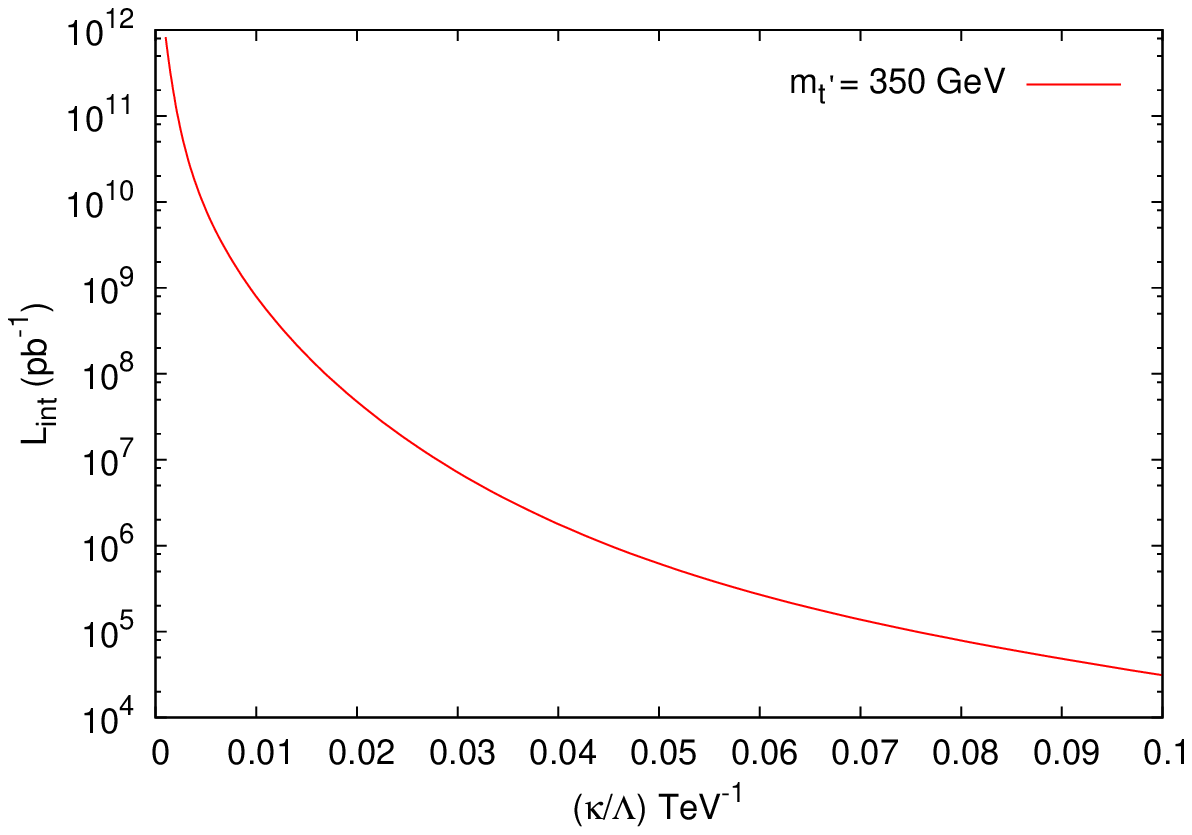}} \hspace{0.1cm}
\subfigure[]{\includegraphics[width=8cm]{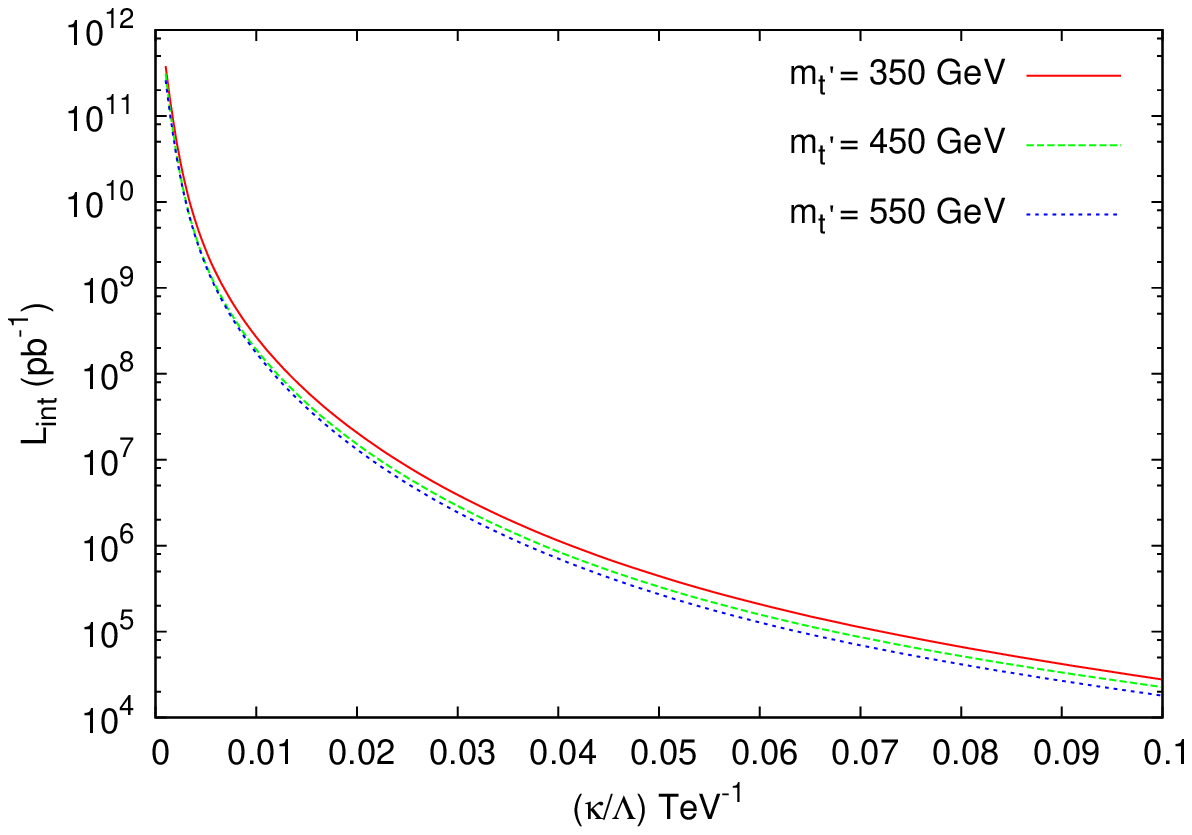}} \caption{The
attainable integrated luminosity for 3$\sigma$ observation limit
depending on anomalous couplings at (a) $\sqrt{s}=$0.5 TeV (b)
$\sqrt{s}=$3 TeV.}\label{lt}
\end{figure}

Finally, in Fig.~\ref{lt}, we plot the lowest necessary luminosities
with 3$\sigma$ observation limits depending on anomalous couplings
at (a) $\sqrt s=0.5$ TeV and (b) 3 TeV energies, respectively. It is
seen that the fourth generation $t'$ quarks with masses 350 GeV can
be observed at 3$\sigma$ observation limit with lowest integrated
luminosity about 3$\times 10^4$ pb$^{-1}$ at ILC and CLIC.

\section{Conclusion}
We find the discovery regions of the parameter space for the single
production of fourth generation $t'$ quarks via anomalous
interaction vertices at the ILC and CLIC energies. If the fourth
generation $t'$ quarks have anomalous couplings that dominate over
the SM chiral interactions they can be produced with large numbers.
Our results shows that, the lower limit of the anomalous couplings
$\kappa_{\gamma}/\Lambda$ and $\kappa_Z/\Lambda$ are found down to
0.07 TeV$^{-1}$ and 0.05 TeV$^{-1}$ for ILC and CLIC, respectively,
assuming a maximal parametrization for extended CKM elements. We
also find the lowest necessary luminosity limit values for the
$e^+e^-$ colliders which will provide a unique opportunity to search
for anomalous couplings of the fourth generation $t'$ quarks.


\begin{thebibliography}{99}
\bibitem{He:2001tp}
  H.~J.~He, N.~Polonsky and S.~f.~Su,
  Phys.\ Rev.\  D {\bf 64}, 053004 (2001).
\bibitem{Kribs:2007nz}
  G.~D.~Kribs, T.~Plehn, M.~Spannowsky and T.~M.~P.~Tait,
  Phys.\ Rev.\  D {\bf 76}, 075016 (2007).

\bibitem{Hou:2010}
  W.~S.~Hou and C.~Y.~Ma,
  Phys.\ Rev.\  D {\bf 82}, 036002 (2010).
  \bibitem{BarShalom:2009sk}
  S.~Bar-Shalom, D.~Oaknin and A.~Soni,
  Phys.\ Rev.\  D {\bf 80}, 015011 (2009).

\bibitem{Buras:2010pi}
  A.~J.~Buras, B.~Duling, T.~Feldmann, T.~Heidsieck, C.~Promberger and S.~Recksiegel,
  JHEP {\bf 1009}, 106 (2010).
\bibitem{Soni:2008bc}
  A.~Soni, A.~K.~Alok, A.~Giri, R.~Mohanta and S.~Nandi,
  Phys.\ Lett.\  B {\bf 683}, 302 (2010).
\bibitem{Eberhardt:2010bm}
  O.~Eberhardt, A.~Lenz and J.~Rohrwild,
  Phys.\ Rev.\  D {\bf 82}, 095006 (2010).

\bibitem{Soni:2010xh}
A.~Soni, A.~K.~Alok, A.~Giri, R.~Mohanta and S.~Nandi,
Phys.\ Rev.\  D {\bf 82}, 033009 (2010).


\bibitem{Alok:2010zj}
 A.~K.~Alok, A.~Dighe and D.~London,
 Phys.\ Rev.\  D {\bf 83}, 073008 (2011).





  \bibitem{Holdom:1986rn}
  B.~Holdom,
  Phys.\ Rev.\ Lett.\  {\bf 57}, 2496 (1986)
  [Erratum-ibid.\  {\bf 58}, 177 (1987)].
\bibitem{Hill:1990ge}
  C.~T.~Hill, M.~A.~Luty and E.~A.~Paschos,
  Phys.\ Rev.\  D {\bf 43}, 3011 (1991).
\bibitem{Elliott:1992xg}
  T.~Elliott and S.~F.~King,
  Phys.\ Lett.\  B {\bf 283}, 371 (1992).
\bibitem{Hung:2010xh}
  P.~Q.~Hung and C.~Xiong,
  Nucl.\ Phys.\  B {\bf 848}, 288 (2011).


\bibitem{Holdom:2006mr}
  B.~Holdom,
  JHEP {\bf 0608}, 076 (2006).
\bibitem{Hung:2007ak}
  P.~Q.~Hung and M.~Sher,
  Phys.\ Rev.\  D {\bf 77}, 037302 (2008).

\bibitem{Hung:2009ia}
  P.~Q.~Hung, C.~Xiong,
  Phys.\ Lett.\  {\bf B694}, 430-434 (2011).
\bibitem{Hung:2009hy}
  P.~Q.~Hung and C.~Xiong,
  Nucl.\ Phys.\  B {\bf 847}, 160 (2011).

\bibitem{Cakir:2009xi}
  O.~Cakir, A.~Senol and A.~T.~Tasci,
  Europhys.\ Lett.\  {\bf 88}, 11002 (2009).
\bibitem{Holdom:2009rf}
  B.~Holdom, W.~S.~Hou, T.~Hurth, M.~L.~Mangano, S.~Sultansoy and G.~Unel,
  PMC Phys.\  A {\bf 3}, 4 (2009).
  %
  \bibitem{Lister:2008is}
  A.~Lister  [CDF Collaboration],
  arXiv:0810.3349 [hep-ex].
 \bibitem{Aaltonen:2009nr}
  T.~Aaltonen {\it et al.}  [CDF Collaboration],
  Phys.\ Rev.\ Lett.\  {\bf 104}, 091801 (2010).
\bibitem{Chanowitz:1978mv}
  M.~S.~Chanowitz, M.~A.~Furman and I.~Hinchliffe,
  Nucl.\ Phys.\  B {\bf 153}, 402 (1979).
\bibitem{atlas}
ATLAS "Detector and Physics Performance Technical Design Report,"
Vol. 2, p. 519 Reports No. CERN-LHCC-99-15 and No. ATLAS-TDR-15, May
1999.
\bibitem{:2007sg}
  J.~Brau, (Ed.) {\it et al.} [ ILC Collaboration ],
  ``ILC Reference Design Report: ILC Global Design Effort and World Wide Study,''
[arXiv:0712.1950 [physics.acc-ph]].
  \bibitem{Braun:2008zzb}
  H.~Braun {\it et al.}, CLIC-NOTE-764, [CLIC Study Team Collaboration], CLIC 2008
  parameters.


\bibitem{Arik:2003vn}
  E.~Arik, O.~Cakir, S.~Sultansoy,
  Europhys.\ Lett.\  {\bf 62}, 332-335 (2003).

\bibitem{Arik:2002sg}
  E.~Arik, O.~Cakir, S.~Sultansoy,
  Phys.\ Rev.\  {\bf D67}, 035002 (2003).

\bibitem{Cakir:2009ib}
  I.~T.~Cakir, H.~Duran Yildiz, O.~Cakir {\it et al.},
  Phys.\ Rev.\  {\bf D80}, 095009 (2009).

\bibitem{Ciftci:2008tc}
  R.~Ciftci,
  Phys.\ Rev.\  {\bf D78}, 075018 (2008).

\bibitem{Sahin:2010wg}
  M.~Sahin, S.~Sultansoy, S.~Turkoz,
  Phys.\ Rev.\  {\bf D82}, 051503 (2010).
\bibitem{Alan:2003za}
  A.~T.~Alan, A.~Senol, O.~Cakir,
  Europhys.\ Lett.\  {\bf 66}, 657-660 (2004).
\bibitem{Ciftci:2009it}
  R.~Ciftci and A.~K.~Ciftci,
  arXiv:0904.4489 [hep-ph].

\bibitem{Pukhov:1999gg} A.~Pukhov \textit{et al.}, 
 arXiv:hep-ph/9908288.

\bibitem{Hou:2006mx}
  W.~S.~Y.~Hou, M.~Nagashima and A.~Soddu,
  Phys.\ Rev.\  D {\bf 76}, 016004 (2007).
%
\bibitem{Fritzsch:1999rd}
  H.~Fritzsch and D.~Holtmannspotter,
  Phys.\ Lett.\  B {\bf 457}, 186 (1999).
  %
\bibitem{Ball:2007zza}
  G.~L.~Bayatian {\it et al.}  [CMS Collaboration],
  J.\ Phys.\ G {\bf 34}, 995 (2007).
  \end{thebibliography}
\end{document}